\documentclass[galaxies,article,accept,pdftex,moreauthors]{Definitions/mdpi} 

\usepackage{newunicodechar}
\newunicodechar{″}{\ensuremath{''}}

\def\mnras{Mon. Not. R. Astron. Soc.}             % Monthly Notices of the RAS

\def\aj{Astron. J.}                   % Astronomical Journal
\def\apj{ Astrophys. J.}                 % Astrophysical Journal
\def\aap{Astron. Astrophys.}                % Astronomy and Astrophysics
\def\jcap{J. Cosmology Astropart. Phys.}
\def\ssr{Space~Sci.~Rev.}     % Space Science Reviews
\def\ssr{Space~Sci.~Rev.}     % Space Science Reviews
\usepackage{xspace}
\usepackage{xcolor}
\firstpage{1} 
\pubvolume{1}
\issuenum{1}
\articlenumber{0}
\pubyear{2026}
\copyrightyear{2026}
\externaleditor{Firstname Lastname} % More than 1 editor, please add `` and '' before the last editor name
\datereceived{4 June 2026} 
\daterevised{21 September 2026} % Comment out if no revised date
\dateaccepted{24 September 2026} 
\datepublished{} 
\Title{Improving Lens Modelling from Ground-Based Imaging \linebreak  with Deconvolution}

\newcommand{\arcs}{\ensuremath{^{\prime \prime}}}

\newcommand{\gext}{\ensuremath{\gamma_{\rm ext}}\xspace}
\newcommand{\thetae}{\ensuremath{\theta_{\rm E}}\xspace}
\newcommand{\nsub}{\ensuremath{n_{\rm sub}}\xspace}
\newcommand{\lhf}{\ensuremath{\lambda_{\rm hf}}\xspace}
\newcommand{\lscale}{\ensuremath{\lambda_{\rm scales}}\xspace}
\newcommand{\qm}{\ensuremath{q_{\rm m}}\xspace}

\Author{Eric Paic %MDPI: Please carefully check the accuracy of names and affiliations.
 $^{1,*}$%MDPI: We added the * for the corresponding author, please confirm.
\orcidA, Kenneth C. Wong $^{1}$\orcidB, Anupreeta More $^{2,3}$\orcidC~and Anton T. Jaelani $^{4,5}$\orcidD }

\AuthorNames{Eric Paic, Kenneth C. Wong, Anupreeta More, Anton T. Jaelani}

\address{%
$^{1}$ \quad Research Center for the Early Universe, Graduate School of Science, The University of Tokyo, 7-3-1 Hongo, Bunkyo-ku, 
 Tokyo 113-0033, Japan; kcwong19@gmail.com
\\
$^{2}$ \quad  Inter-University Centre for Astronomy and Astrophysics (IUCAA),
 Post Bag 4, Ganeshkhind, \linebreak  Pune 411 007
, India; anupreeta@iucaa.in 
 \\
$^{3}$ \quad Kavli IPMU (WPI), UTokyo Institutes for Advanced Study (UTIAS), 
 The University of Tokyo, Kashiwa
 277-8583, Japan \\
$^{4}$ \quad Astronomy Research Group, Fakultas Matematika dan Ilmu Pengetahuan Alam (FMIPA) 
 Institut Teknologi Bandung, Jl. Ganesha 10, Bandung 40132, Indonesia; antontj@itb.ac.id
 \\
$^{5}$ \quad University Center of Excellence for Space Science, Technology, and Innovation, Institut Teknologi Bandung, Jl. Ganesha 10, Bandung 40132, Indonesia \\
}

\corres{Correspondence: eric.paic@gmail.com 
}

\abstract{Accurate and precise mass modelling of strong lensing systems is essential for extracting insights into cosmology, galaxy environments, and the nature of dark matter. The Hyper Suprime-Cam Subaru Strategic Program (HSC-SSP), which has mapped $\sim$1000 square degrees of the sky, discovered over 1000 definite or probable galaxy-scale gravitational lens candidates. As the resolution of ground-based imaging poses challenges for accurate lens modelling, high-resolution imaging is traditionally  sought after for detailed analyses of these systems. In this work, we instead propose applying the STARRED algorithm, which leverages starlet wavelet regularisation, as a preprocessing step before lens modelling to maximise the scientific output of the large HSC-SSP sample. We test this approach on seven HSC-like mock lensing systems with different lensing configurations. Our results show that the resolution boost from $\sim$0.6\arcs~to $\sim$0.15\arcs~enables improves accuracy and sub-3\% precision measurements of key lensing parameters such as the Einstein radius, \thetae, and the mass axis ratio, \qm, for ground-based images. These results pave the way for accurate measurements of key strong lensing quantities from a large sample of ground-based observations.}

\keyword{\textls[-25]{galaxy--galaxy lensing; strong lensing mass modelling; deconvolution; ground-based imaging; SuGOHI; HSC-SSP; STARRED; Herculens}}

\begin{document}

%%%%%%%%%%%%%%%%%%%%%%%%%%%%%%%%%%%%%%%%%%

\section{Introduction}
Strong gravitational lensing is a manifestation of Einstein’s general relativity, where the gravitational field of a massive object, such as a galaxy or galaxy cluster, bends the light from a background source, producing multiple distorted images, arcs, or Einstein rings \linebreak  (see \cite{saha24}, for detailed formalism). When the lens is an isolated galaxy, this phenomenon enables a diverse range of astrophysical and cosmological investigations, such as constraining the nature of dark matter via subhalo detection (see \cite{vegetti24} for a complete review), studying the universality of the stellar initial mass function (e.g., \cite{treu10, shajib24}), understanding AGN structures (e.g., \cite{paic22, sluse24}) and measuring the expansion rate of the Universe (e.g., \cite{birrer24, tdcosmo25, paic26}).

The accuracy of these studies hinges critically on the precision of the lens mass model. A robust mass model is essential for reconstructing the source features, constraining cosmology from measured time delays, and inferring the distribution of dark matter. However, constructing models with $\sim$1\% precision for parameters such as the Einstein radius \thetae traditionally requires high-resolution imaging data, typically obtained from space-based telescopes such as the Hubble Space Telescope (HST) and the James Webb Space Telescope (JWST). While both continue to provide exquisite data for a subset of known lenses, their limited observing time and field of view mean that only a small fraction of the thousands of strong lenses discovered in recent years have been imaged at the required resolution.
The explosion in the number of known strong lensing systems over the past decade is largely attributable to wide-field surveys such as Gaia (e.g., \cite{lemon18, lemon19}) and the Hyper Suprime-Cam Subaru Strategic Program (HSC-SSP) (e.g., \cite{jaelani21,jaelani24}). We recommend \cite{lemon24} for a full review of lens-finding techniques. %Please verify.
These surveys have identified thousands of new lens systems, but the majority lack high-resolution follow-up observations. To address this gap, many studies have relied on ground-based imaging, which, while more accessible, suffers from atmospheric turbulence and lower spatial resolution. The resulting point spread function (PSF) blurs the lensed images, which increases blending of the lens and source light. This effect degrades the precision of the mass models and limits the scientific \linebreak  return (e.g., \cite{knabel23, poh25}).
To mitigate these challenges, we propose a novel approach that deconvolves the ground-based observations. %Please verify.
Deconvolution aims to reverse the blurring effects of the PSF, thereby recovering higher-resolution information from ground-based imaging. By enhancing the resolution of ground-based data, we aim to improve the accuracy of lens mass models and unlock the full potential of the vast number of strong lenses discovered in wide-field surveys. The goal of this paper is to validate such methods to facilitate accurate modelling of HSC lenses gathered in the Survey of Gravitationally Lensed Objects in HSC Imaging (SuGOHI) project
 (Accessed \href{https://www-utap.phys.s.u-tokyo.ac.jp/~oguri/sugohi/}{here} on 28.09.2026). In Section \ref{sec:methods}, we present the deconvolution and lens modelling methods used and the HSC-like mock systems we use to benchmark our method. In \mbox{Section \ref{sec:results}}, we demonstrate the efficacy of this method and its implications for future lensing studies using ground-based images. We discuss future applications of this method and conclude the paper in Section \ref{sec:conclusion}

%%%%%%%%%%%%%%%%%%%%%%%%%%%%%%%%%%%%%%%%%%
\section{Materials and Methods \label{sec:methods}}

\subsection{HSC-Like Mock Lens Systems}

The HSC-SSP \citep{aihara19} is a wide-field imaging survey conducted using the Hyper Suprime-Cam (HSC, \citep{miyazaki18}), a CCD camera with 0.168\arcs/pix resolution mounted on the 8.2 m Subaru Telescope atop Mauna Kea, Hawaii. The camera is equipped with a set of broadband filters (\textit{g}, \textit{r}, \textit{i}, \textit{z}, and \textit{y}) and narrowband filters, enabling observations across a wide range of wavelengths from the ultraviolet to the near-infrared.
The HSC-SSP survey is designed to reach great depths through total exposure times of approximately 600 s per filter. Mauna Kea's excellent natural seeing conditions enable high spatial resolution, with a median seeing of 0.6\arcs~in the $i$-band. 

For this work, we use simulated galaxy-scale lens systems generated by Ref.~\cite{jaelani24} to test our method. 
We refer to the original work for details about these simulations and summarise the parametrisations used. Following standard assumptions, the mass of the lens is parametrised with an Elliptical Power Law (EPL), defined as:
\begin{equation}  \kappa\left(\theta_{\rm 1},\theta_{\rm 2}\right) = \frac{3-\gamma}{2}\left(\frac{\theta_{\rm E}}{\sqrt{\theta_{\rm 1}^2+\theta_{\rm 2}^2/\qm^2}}\right)^{1-\gamma} ,
\label{eq:SIE}
\end{equation}
with \thetae being the Einstein radius, $\gamma$ the slope, \qm the axis ratio and $\phi$ the position angle defined east of north. By fixing $\gamma=2$, we recover the Singular Isothermal Elliptical (SIE) profile. 
The aggregate effect of line-of-sight adjacent galaxies is added through external shear:
\begin{equation}
    \vec{\gamma_{\rm{ext}}} = \gamma_{\rm{ext}} \begin{pmatrix} \cos 2\psi & \sin 2\psi \\ \sin 2\psi & -\cos 2\psi \end{pmatrix},
    \label{eq:shear}
\end{equation}
with \gext being the magnitude and $\psi$ its position angle.

The Sersic profile \cite{sersic63} is used to parametrise the light components of the simulation:
\begin{equation}
    I\left(\theta_{\rm 1},\theta_{\rm 2}\right)=I_{\rm eff}\exp \left\{-b_{ n}\left[\left(\frac{\sqrt{q_{\rm l}\theta_{\rm 1}^2+\theta_{\rm 2}^2/q_{\rm l}}}{R_{\rm Sersic}}\right)^{ 1/n}-1\right]\right\},
 \label{eq:sersic}
\end{equation}
where $b_{ n}$ is the normalising factor; $R_{ \rm Sersic}$ is the effective radius (the product average of the semi-major and semi-minor axes), also defined as the half-light radius; $I_{ \rm eff}$ is the intensity at $R_{ \rm Sersic}$; $q_{\rm l}$ is its axis ratio; and $n$ is the Sersic index. 

\begin{figure}[H]
%\isPreprints{\centering}{} % Only used for preprints
\includegraphics[width=0.9\textwidth]{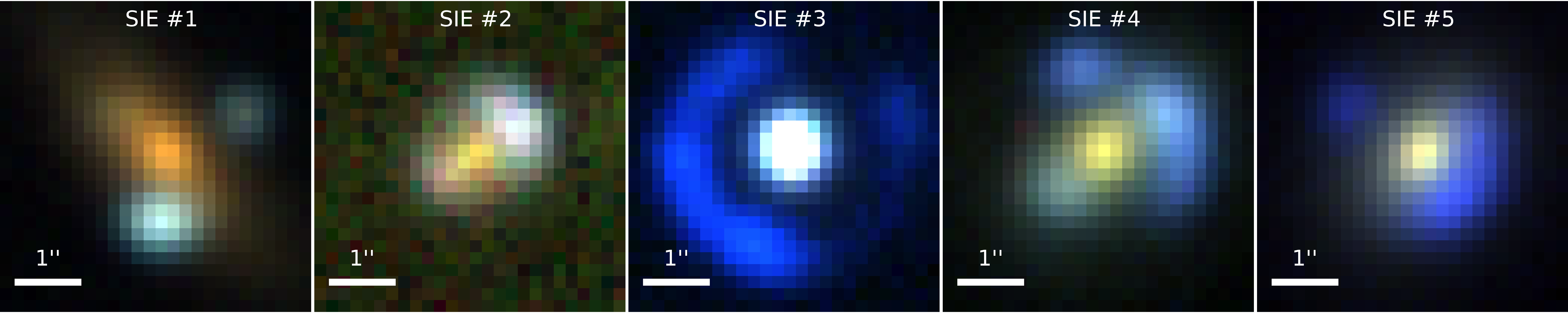}

\includegraphics[width=0.36\textwidth]{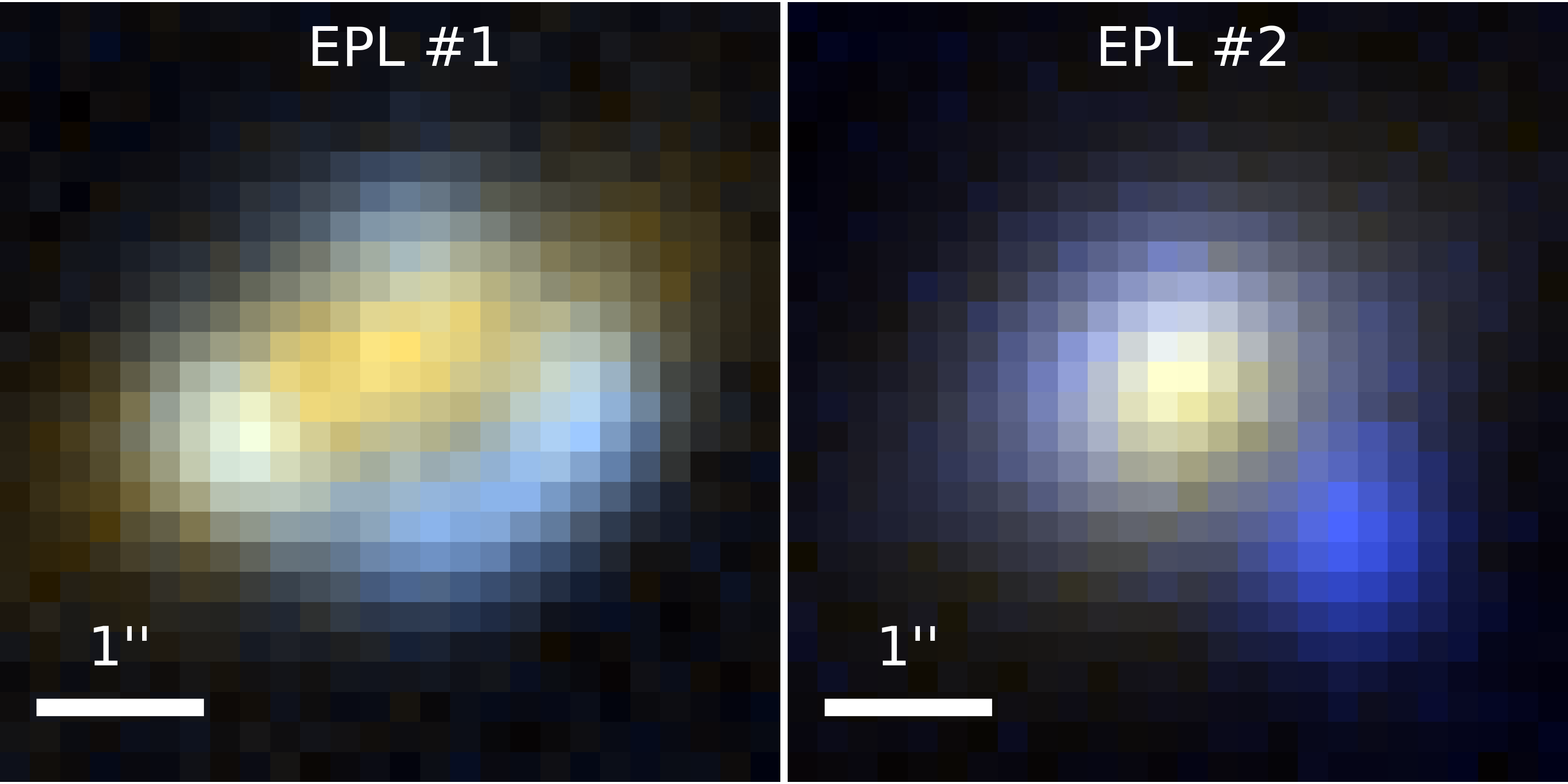}
\caption{Composite
 image of the 5 SIE and 2 EPL systems used, made with filters \textit{g}, \textit{r} and \textit{i}. SIE \#1 is a highly elliptical fold lens with one image blended with the lens light; SIE \#2 is a small-separation double; SIEs \#3, \#4, and \#5 are quadruply lensed cusps with variable source separation and luminosity. EPLs \#1 and \#2 are quadruply lensed cusps with extreme power-law indices: steep ($\gamma=2.2$) and shallow ($\gamma=1.87$), respectively. \label{fig:rgb}}
\end{figure}

In most known lens galaxies, the mass profile is compatible with an SIE \linebreak  profile (e.g., \cite{shajib19,schmidt23}), and so we focus our study on 5 mocks created with SIE profiles. However, to account for systems deviating from an SIE profile and avoid circularity bias of the study, we also test our approach on 2 EPL mocks %Please verify.
with extremely shallow and steep slopes of $\gamma=1.87$ and $\gamma=2.20$. 
The parameters for each of these profiles are drawn from the wide ranges given in Table 1 of Ref.~\cite{jaelani24}, and Figure~\ref{fig:rgb} displays the 5 SIE and 2 EPL mock systems used in our experiment. To ensure robustness and applicability to the diverse range of lensing configurations expected in large surveys, we selected mock systems spanning a variety of separations and morphologies, including challenging cases such as highly elliptical lenses (SIE \#1) and small separations (SIE \#2). This approach allows us to validate the method's reliability across the full spectrum of realistic scenarios, ensuring its effectiveness when applied to the broader population of lenses discovered in wide-field surveys. 
Although our sample is not a fully controlled parameter grid, several
pairs of mocks isolate the effect of individual lens properties: SIEs
\#4 and \#5 share a similar \thetae but differ in ellipticity; SIEs
\#1 and \#4 share a similar \thetae and ellipticity but differ in
lensing configuration; and SIEs \#2 and \#3 share a similar ellipticity
but differ in \thetae. These pairs allow us to partially disentangle
the impact of \thetae, ellipticity, and configuration on the
performance of the method, in addition to the broader diversity
spanned by the full sample.

An additional selection criterion for the mocks is the visibility of the source in at least two bands. In our case, the bands \textit{r} and \textit{g} provide the best signal-to-noise ratio (SNR) for the source light, which is the main source of constraint for lens mass modelling. We will therefore conduct our analysis by simultaneously modelling these two bands.

\subsection{Deconvolution Formalism} %MDPI: This heading is duplicated with Section 3.1, please check if both of them should be retained.

The STARRED package \cite{michalewicz23} leverages isotropic starlet wavelets \cite{starck16}, an optimal basis for representing astronomical objects, to regularise deconvolution solutions. STARRED employs a dual-channel approach, separating point sources from extended sources, and avoids full PSF removal. Instead, it enhances resolution to a known Gaussian PSF, reducing deconvolution artefacts that arise from targeting infinite resolution, which is a common pitfall in other methods. Additionally, it uses {\tt JAX} automatic differentiation (Accessed \href{https://github.com/jax-ml/jax}{here} on 28.09.2026) %replaced the citation by the URL of the github to find the package
 to optimise the high-dimensional problem efficiently and more quickly than traditional methods.

We refer to \cite{millon24} for a detailed development of the algorithm and provide a summary here. Let $d(x)$ be an observation corrupted by additive noise $n(x)$. The goal of the deconvolution is to recover $f(x)$ in the following formulation of $d(x)$:
\begin{equation}
    d(x) = s(x) \ast f(x) + n(x),
\end{equation}
where $s(x)$ is a narrow function.
Here, $f$ is represented as a function of free parameters $a_k$ and $c_k$ such that
\begin{equation}
f(x) = h(x) + \sum_{k=1}^{M} a_k r(x - c_k),   
\label{eq:resolution}
\end{equation}
where $r(\mathbf{x})$ is a 2-pixel Full Width at Half-Maximum (\text{FWHM}) Gaussian; $h(\mathbf{x})$ is a grid of pixels; $M$ is the number of point sources located at $\mathbf{c}_k = \begin{bmatrix} c_{x_k} & c_{y_k} \end{bmatrix}^\top$ for $k = 1, \dots, M$; and $a_k$ represents the Gaussian amplitudes associated with each point source.%Please verify.

%With $r(\mathbf{x})$ a 2-pixel Full Width Half Maximum (FWHM) Gaussian, $h(\mathbf{x})$ a grid of pixels, and $M$ the number of point sources located at $\mathbf{c}k = \begin{bmatrix} c{x_k} & c_{y_k} \end{bmatrix}^\rm{T}$ for k = 1, ..., M, with $a_k$ being the Gaussian amplitudes associated with each point source.

The objective function to be minimised is

\begin{adjustwidth}{-\extralength}{0cm}
%\centering %% If there is a figure in wide page, please release command \centering
\begin{equation}
    S = \sum_{i=1}^{N_e} \frac{1}{\sigma_i^2} \left[ \left( s_i \ast h_i + s_i \ast \sum_{k=1}^{M} a_k r(x_i - c_k) \right) \downarrow - d_i \right]^2 + \lscale \phi_{\rm{scales}}(h(x))_1 + \lhf \phi_{\rm{hf}}(h(x))_1,
    \label{eq:dec}
\end{equation}
\end{adjustwidth}
where $N_e$ is the number of epochs, $\phi(\cdot)$ is the starlet transform operator, $\downarrow$ denotes the downsampling operation, and $\sigma_i$ is the noise level at each pixel of the data. The parameters \lhf and \lscale are Lagrange multipliers that weight the highest frequencies of the pixelated model component $h(x)$ and all other scales (except the coarsest one), respectively.

For this work, we consider galaxy--galaxy strong lenses (i.e., $M=0$ and $N_e=1$, since there is no luminosity variability); hence, the main hyperparameters at hand are \lscale, \lhf, and the subsampling factor \nsub.

As \lscale is parametrised in units of noise, it acts as a soft threshold for selecting the features to reconstruct. In other words, any feature more than \lscale-$\sigma$ above the noise %Please verify.
will be accounted for by the regularised solution $h(x)$.

We compute the noise map of the deconvolved cut-out through a
Fisher-information analysis of the deconvolution objective $S$
(Equation~(\ref{eq:dec})), evaluated at the best-fit pixel grid $h(x)$. This analysis yields the full
pixel--pixel noise covariance matrix of the deconvolved cut-out. Deconvolution correlates noise spatially between neighbouring pixels, but, to minimise the computational cost, for the lens-modelling likelihood used throughout this work, we adopt its diagonal, i.e., a per-pixel $\sigma_i$. In Appendix~\ref{app:noisemap}, we investigate the impact of this diagonal approximation by repeating the lens-modelling fit with the full covariance matrix propagated into a correlated-noise likelihood, and we quantify the resulting change in precision and accuracy on \thetae, \qm, and $\gamma_{\rm ext}$.

\subsection{Lens Modelling}
The lens modelling package {\tt Herculens} \cite{galan2022}, also based on JAX, is computationally efficient, making it suitable for modelling large samples of lensing systems.
Both STARRED and Herculens leverage JAX's automatic differentiation \citep{jax18} to accelerate the optimisation of high-dimensional models. As an indication of the resulting performance, deconvolution of a single cut-out with STARRED takes $\sim$5 min per band, and lens modelling of a single system with Herculens (optimisation and SVI across 5 starting points) takes \mbox{$\sim$1 h} for all 5 starting points combined, both on an Intel(R) Core(TM) Ultra 9 185H CPU. For comparison, a similar fit using a non-JAX-based pipeline (\texttt{lenstronomy}, Birrer and \linebreak  Amara \cite{birrer18}, Birrer et al. \cite{birrer21}) takes $\sim$10~h on the same hardware.

A common choice for lens modelling with low-resolution data is to model the lens mass distribution using an SIE and external shear as defined in Equations~(\ref{eq:SIE}) and (\ref{eq:shear}).

When not masked, the lens light is modelled with a combination of two Sersic profiles (see Equation~(\ref{eq:sersic})) with $n=1$ and $n=4$ with shared centroids, while the source is modelled with a Sersic profile (with a free $n$).

The PSF and its noise map for the HSC-like data are provided by the HSC pipeline tool for each band. For the deconvolved data, the PSF is fixed by construction; hence, we model the PSF as a Gaussian with an FWHM of 2 pixels, i.e., the width of the narrow function $f(x)$ retrieved by the deconvolution in Equation (\ref{eq:resolution}). 

For the modelling workflow, we initialise 40 starting points within conservative priors across the parameter space. The priors used for the source light profile ellipticity and position angles are the same as the ones used to generate the simulated systems in \cite{jaelani24}. For the mass model, we estimate \thetae as the separation between the most luminous point and the centre of the lens, and we use a uniform prior with a $\pm$0.2$^{\prime\prime}$ range around this estimate. If the lens exhibits clear ellipticity (e.g., mocks \#1 and \#4), we perform a preliminary fit of the lens light and use, for \qm, a uniform prior with a range of $\pm$0.3 around the measured lens light axis ratio $q_{\rm l}$. Otherwise, we adopt a default uniform prior between \mbox{0.4 and 1}. The specific priors for the mass parameters for each mock are given in Table~\ref{tab:priors}. The position angle of the mass distribution, $\phi$, and the
orientation of the external shear, $\psi$, are also free parameters
in the fit, with uniform priors of
$[-180^\circ,180^\circ]$ for both angles.
The light-profile parameters ($I_{\rm eff}$,
$R_{\rm Sersic}$, $q_l$, and $n$ for the source and lens Sersic
components) are likewise free, with broad uninformative priors. During the fitting procedure, the mass profile parameters are optimised jointly across the different bands while the source and lens light parameters are constrained separately.
 We use {\tt optax}'s AdaBelief optimiser \cite{zhuang20} via a wrapper implemented in Herculens to identify the 5 starting points for which the optimised models yield the highest likelihood. We then use {\tt Numpyro}'s \cite{phan19} Stochastic Variational Inference (SVI) with a full-covariance Gaussian approximation to sample the posterior of the lens model using the predetermined starting points. While the best starting point from the optimisation typically yields the highest likelihood with SVI, the stochastic nature of SVI means this is not guaranteed. To account for this variability, we conservatively sample the posterior using all 5 starting points. For a given system, the final posterior is the likelihood-weighted combination of the obtained posteriors.

\begin{table}[H] 
\caption{Priors
 used for the mass model parameters and the corresponding true value. For \gext, the same prior is used for every system. \label{tab:priors}}
%\newcolumntype{C}{>{\centering\arraybackslash}X}

\renewcommand{\arraystretch}{1.2}
\renewcommand{\aboverulesep}{.1pt}
\renewcommand{\belowrulesep}{.1pt}

\begin{adjustwidth}{-\extralength}{0cm}
%\centering %% If there is a figure in wide page, please release command \centering
\begin{minipage}{\fulllength}
\begin{tabularx}{\textwidth}{CCCCCCCC}
\toprule
\textbf{Prior} & \multirow{2}{*}{\textbf{SIE \#1}} & \multirow{2}{*}{\textbf{SIE \#2}} & \multirow{2}{*}{\textbf{SIE \#3}} & \multirow{2}{*}{\textbf{SIE \#4}} & \multirow{2}{*}{\textbf{SIE \#5}} & \multirow{2}{*}{\textbf{EPL \#1}}& \multirow{2}{*}{\textbf{EPL \#2}}\\
\textbf{Truth} &  & &  && & \\
\midrule
\multirow{2}{*}{\thetae [\arcs]} & [0.8, 1.2] & [0.5, 0.9] &[1.3, 1.7]& [0.8, 1.2] & [0.8, 1.2]& [0.8, 1.2]& [0.5, 0.9] \\
 & 0.93& 0.67 & 1.51 & 1.02 & 0.99 & 0.94 & 0.77 \\
 \midrule
\multirow{2}{*}{\qm} & [0.1, 0.7] & [0.4, 1] & [0.4, 1] & [0.2, 0.8] & [0.4, 1]& [0.4, 1]& [0.4, 1] \\
 & 0.51 & 0.86 & 0.88 & 0.62 & 0.74 & 0.96 & 0.95 \\
\midrule
\multirow{2}{*}{\gext} & [0.0001, 0.2]& [0.0001, 0.2]& [0.0001, 0.2]& [0.0001, 0.2]& [0.0001, 0.2]& [0.0001, 0.2]& [0.0001, 0.2] \\
 & 0.008 & 0.02 & 0.01 & 0.02 & 0.02& 0.15 & 0.12 \\

 \bottomrule
\end{tabularx}
\end{minipage}
\end{adjustwidth}
\end{table}

%\begin{table}[H]\ContinuedFloat
%
%\caption{{\em Cont.}}
%\renewcommand{\arraystretch}{1.2}
%\renewcommand{\aboverulesep}{.1pt}
%\renewcommand{\belowrulesep}{.1pt}

%\begin{adjustwidth}{-\extralength}{0cm}
%%\centering %% If there is a figure in wide page, please release command \centering
%\begin{minipage}{\fulllength}
%\begin{tabularx}{\textwidth}{CCCCCCCC}
%\toprule
%\textbf{Prior} & \multirow{2}{*}{\textbf{SIE \#1}} & \multirow{2}{*}{\textbf{SIE \#2}} & \multirow{2}{*}{\textbf{SIE \#3}} & \multirow{2}{*}{\textbf{SIE \#4}} & \multirow{2}{*}{\textbf{SIE \#5}} & \multirow{2}{*}{\textbf{EPL \#1}}& \multirow{2}{*}{\textbf{EPL \#2}}\\
%\textbf{Truth} &  & &  && & \\

%\end{tabularx}
%\end{minipage}
%\end{adjustwidth}
%\end{table}

%\vspace{-2cm}
\section{Results \label{sec:results}}

\subsection{Deconvolution}

We conservatively set \lscale = 1, but we need to aggressively denoise the coarsest starlet. Given the clear and distinctive source arc and lens light, SIE \#3 presents an ideal configuration to test the validity of the deconvolution process.

\begin{figure}[H]

\begin{adjustwidth}{-\extralength}{0cm}
\centering %% If there is a figure in wide page, please release command \centering
    \includegraphics[width=1\linewidth]{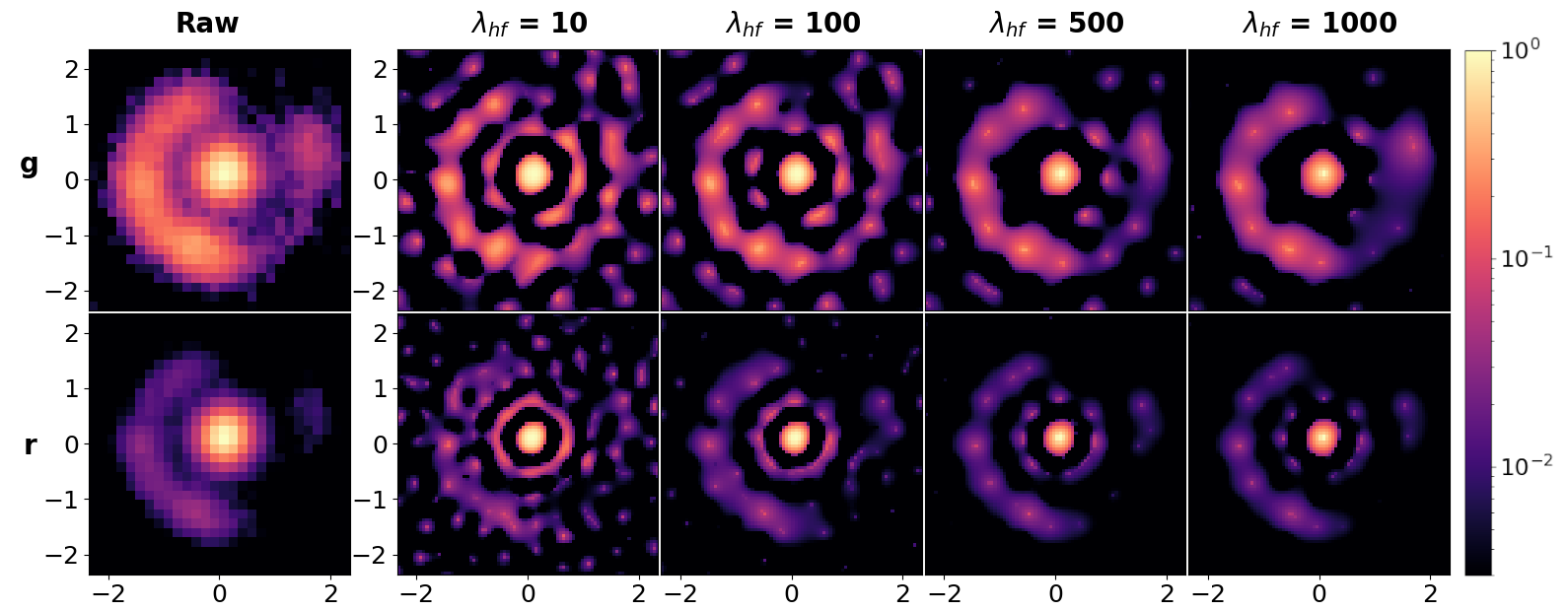}
\end{adjustwidth}
    \caption{Deconvolution
 of SIE \#3 with fixed $\lscale = 1$, \nsub= 3 and $\lhf$ from 10 to 1000.}
    \label{fig:ex_deconv}
\end{figure}

\begin{figure}[H]

\begin{adjustwidth}{-\extralength}{0cm}
\centering %% If there is a figure in wide page, please release command \centering
    \includegraphics[width=\linewidth]{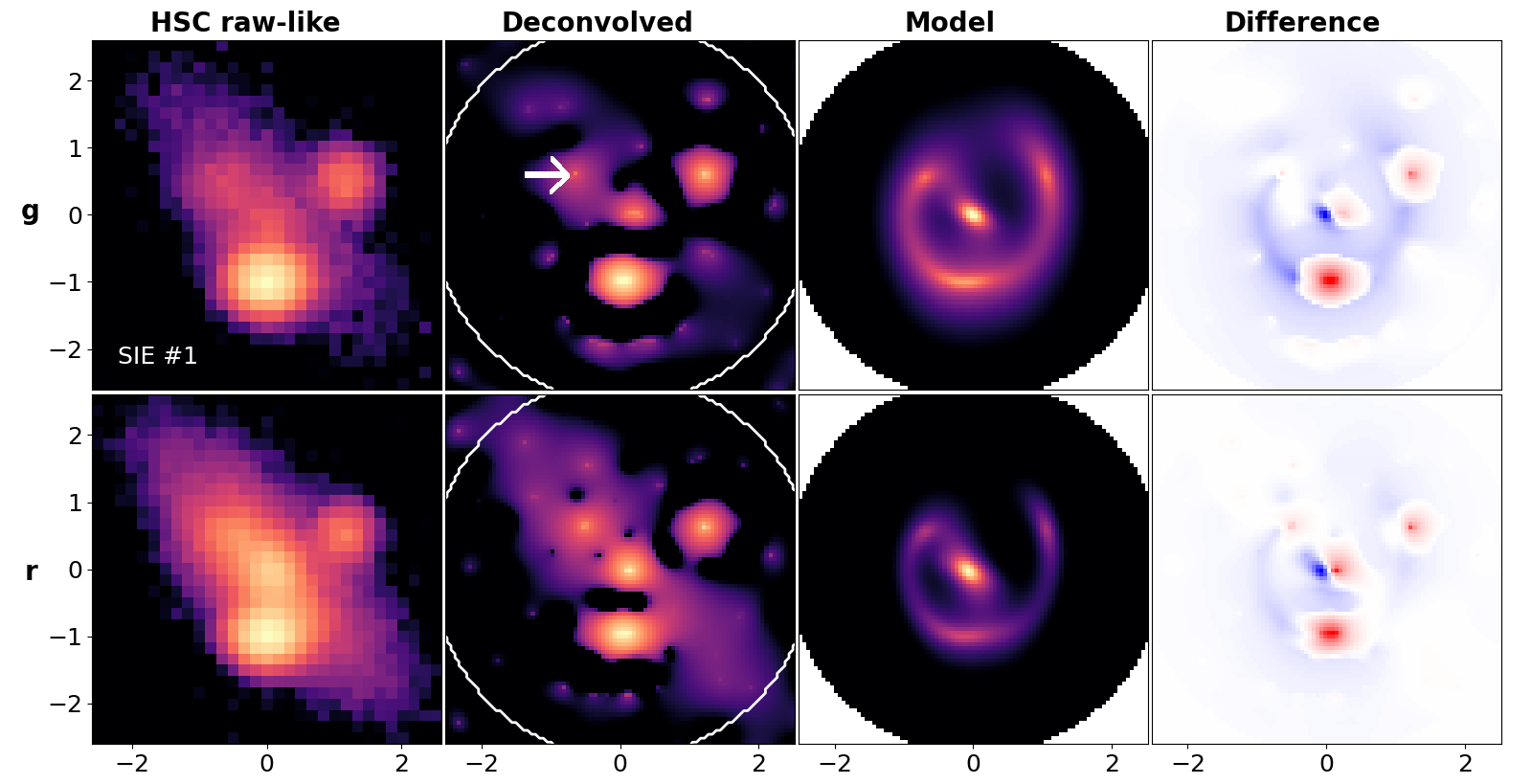}
\end{adjustwidth}
    \caption{Lens modelling of SIE \#1. From left to right: the original HSC data, the deconvolved data, the model obtained with the deconvolved data, and the difference between the latter two. If a mask is used on the deconvolved data, we show its contours in white in the second column and mask the corresponding area in the third and fourth columns. The white arrow in the second column points to the area unveiled by the fourth image.
 }
    \label{fig:modelling_mocks13}
\end{figure}

As shown in Figure~\ref{fig:ex_deconv}, we find that setting \lhf = 1000 is necessary to recover the main features of the lens system and eliminate artefacts outside the source arc that could otherwise affect lens modelling. Increasing \lhf could remove even more of these artefacts, but it would also discard a larger portion of the signal as noise. As we aim to be as conservative as possible, we settle on using
$\lhf = 1000$ for the rest of this work. We verify in Appendix~\ref{app:lhf} that the recovered \thetae and \qm are not sensitive to this choice.
Data with multiple exposures would intrinsically add dithering and hence could allow us to reduce \lhf, ideally matching \lscale = \lhf as in Ref.~\cite{dux25}. The deconvolution of all mocks with \lscale = 1 and \lhf = 1000 is shown in the second column of Figures~\ref{fig:modelling_mocks13} and~\ref{fig:modelling_mocks45}. 

In SIEs \#2, \#3, and \#4 as well as EPL \#2, deconvolution enhances the sharpness and resolution of the source arc. In SIE \#1, it reveals the fourth image (highlighted by the arrow in Figure~\ref{fig:modelling_mocks13}), which was previously blended with the lens light in both bands, while in SIE \#5 and \linebreak  EPL \#1, it separates the source light from the lens light in the \textit{r}-band.

\subsection{Modelling}

\begin{figure}[H]

    \includegraphics[width=0.85\textwidth]{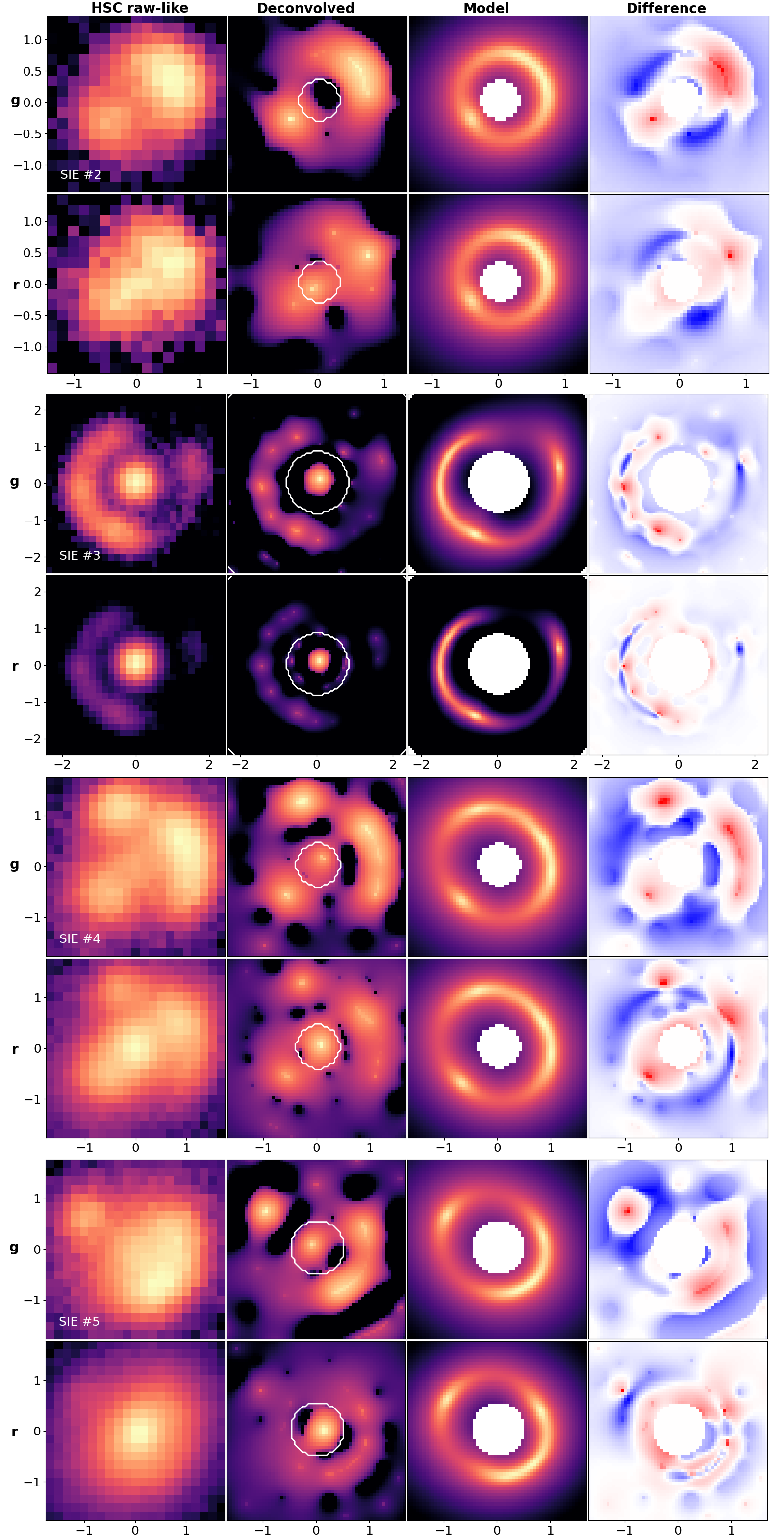}
    \caption{Same as Figure~\ref{fig:modelling_mocks13} for SIEs \#2 to \#5.}
    \label{fig:modelling_mocks45}
\end{figure}

As illustrated in the second column of Figures~\ref{fig:modelling_mocks13} and~\ref{fig:modelling_mocks45}, since deconvolution is data-driven and does not consider any lensing information, the deconvolved source light becomes a series of cuspy dots. While this enhances the arc's sharpness, it prevents the source light from being fitted down to the noise level using a standard Sersic profile. Because the optimisation seeks to match the highest-amplitude peaks left by the data-driven deconvolution, rather than a smooth arc, it tends to converge towards narrower and more elliptical source solutions than the true, smooth source light distribution. This explains why the fitted source appears more compact and elongated than the source light seen in the \mbox{deconvolved cut-outs}.

\begin{figure}[H]
   
\begin{adjustwidth}{-\extralength}{0cm}
%\centering %% If there is a figure in wide page, please release command \centering
 \centering
    \includegraphics[width=0.93\linewidth]{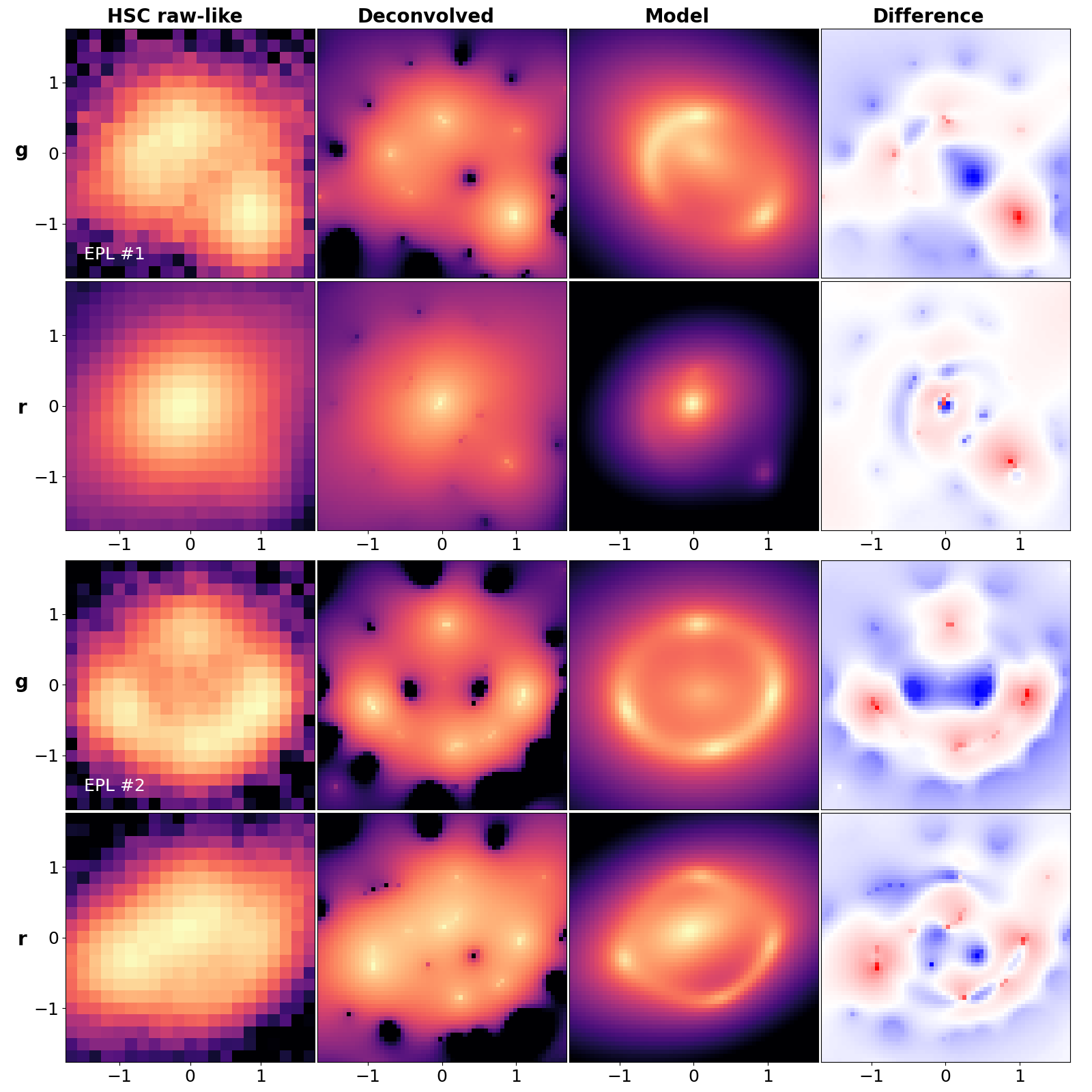}
\end{adjustwidth}
    \caption{Same as Figure~\ref{fig:modelling_mocks13} for EPLs \#1 and \#2.}
    \label{fig:modelling_epl}
\end{figure}

Given that the positions of these dots do not necessarily correspond to consistent locations in the source plane, even more flexible profiles, such as shapelets, fail to accurately model these features. Furthermore, introducing additional degrees of freedom to account for non-physical artefacts could introduce biases into the analysis.
Therefore, as shown in the third column of Figures~\ref{fig:modelling_mocks13}--\ref{fig:modelling_epl}, we focus on reconstructing the overall shape of the arc. The residual images in the difference column confirm that, for every system, the maximum of the source light arc (i.e.,\ the centre of the source) maps to the position of a cuspy dot. Hence, while deconvolution breaks the smoothness of the arc and the overall shape of the source is not perfectly recovered, the position of its brightest features is correctly reproduced.
Consequently, even though the overall shape of the reconstructed source is altered by deconvolution artefacts, the measurements of \thetae and \qm presented in this work \mbox{remain robust}.

\begin{figure}[H]
    
    \includegraphics[width=0.98\textwidth]{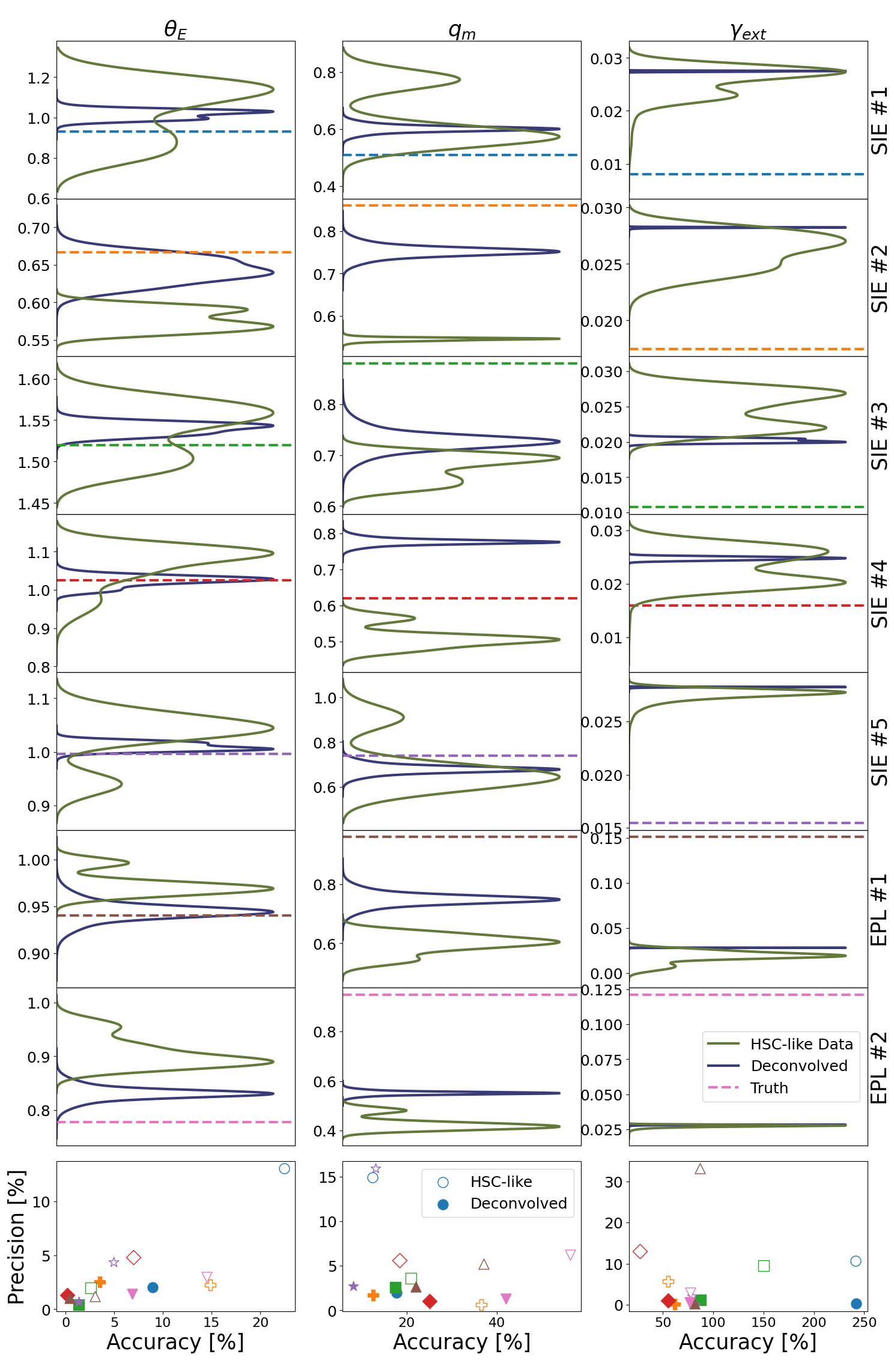}
    \caption{From left to right: \thetae, \qm and \gext measurements for each SIE system. From top to bottom, each row corresponds to a mock system. The bottom row summarises the precision and accuracy of each SIE's measurement for the column's parameter. Each system is represented by a different marker and colour (the same colour used for the ``Truth'' dashed lines in the system's respective row); empty and filled markers, respectively, stand for HSC-like and deconvolved measurements.}
    \label{fig:thetae}
\end{figure}

In Figure~\ref{fig:thetae}, we compare the measurements of \thetae, \qm and \gext obtained with the HSC-like and deconvolved HSC data by computing their precision (standard deviation divided by mean value) and accuracy (difference between the truth and the mean value divided by the latter). For \thetae, we note that modelling the deconvolved data significantly improved both accuracy (by up to six percentage points) in every system except SIE \#4 and precision (by up to seven percentage points) in all systems. The gain in accuracy arises because deconvolution enhances the resolution of the source light and enables efficient masking of the lens light (we do not mask the lens light in SIE \#1, EPL \#1 and EPL \#2 because it still overlaps the source light after deconvolution, especially in the \textit{r}-band, as shown in Figures~\ref{fig:modelling_mocks13} and \ref{fig:modelling_epl}), which is particularly effective for short-separation systems such as SIE \#2. Precision gains are most noticeable in SIEs \#1 and \#5, most likely because deconvolution revealed new constraints that were not available with HSC-like data. 

%\begin{figure}%[H]
%    \centering
%    \includegraphics[width=\linewidth]{compare_allvertical_nomask_jointsource_jointell_subs3_hf100_scales1.0_noiseestTrue_diagonlyTrue_EPL.png}
%    \caption{Same as \ref{fig:thetae} for EPL \#1 and \#2.}
%    \label{fig:thetae_epl}
%\end{figure}

Similar benefits of the deconvolution are observed on the measurement of the lens mass axis ratio, \qm. We note, however, that  there is no gain in accuracy for SIE \#1, which has the most elliptic lens. 
Still, the third column of Figure \ref{fig:thetae} shows that \gext is not recovered, with neither the raw-like nor the deconvolved data reaching an accuracy that would make the measurement usable. This may be due to degeneracies between the external shear (both strength and direction) and the ellipticity of the lens mass (axis ratio and position angle).
We note in addition that the apparently large accuracy values reported
for $\gamma_{\rm ext}$ (up to
$\sim$250\%) partly reflect the small true
magnitude of the external shear in our mocks
($\gamma_{\rm ext} \lesssim 0.02$ in most systems; see Table~\ref{tab:priors}). This metric
sensitivity, compounded with the shear--ellipticity degeneracy noted
above, means that the large accuracy percentages for $\gamma_{\rm ext}$
should not be read as indicating a comparably large absolute error.
We further test whether deconvolution biases the recovered shear
direction $\Psi_{\rm ext}$ towards a fixed value which could bias the modelled \gext. Table~\ref{tab:psiext}
reports $\Psi_{\rm ext}$ measured with the HSC-like and deconvolved
data for every system in our sample; the deconvolved measurements
span $-36^\circ$ to $68^\circ$, with uncertainties overlapping the
HSC-like measurements in every case, ruling out a systematic bias in
a fixed direction.
 
\begin{table}[h]
\caption{$\Psi_{\rm ext}$ measured for every system with HSC-like and
deconvolved data.}
\label{tab:psiext}

\begin{adjustwidth}{-\extralength}{0cm}
%\centering %% If there is a figure in wide page, please release command \centering
\begin{minipage}{\fulllength}
\begin{tabularx}{\textwidth}{LCCCCCCC}
\toprule
 & \textbf{SIE \#1} & \textbf{SIE \#2} & \textbf{SIE \#3} & \textbf{SIE \#4} & \textbf{SIE \#5} & \textbf{EPL \#1} & \textbf{EPL \#2} \\
\midrule
$\Psi_{\rm ext}^{Deconvolved}$ & $68\pm25$ & $22\pm20$ & $-36\pm17$ & $-26\pm21$ & $-22\pm29$ & $22\pm15$ & $-22\pm17$ \\
$\Psi_{\rm ext}^{HSC-like}$    & $15\pm30$ & $23\pm80$ & $1\pm18$   & $18\pm53$  & $-22\pm44$ & $0\pm22$  & $-22\pm30$ \\
\bottomrule
\end{tabularx}
\end{minipage}
\end{adjustwidth}
\end{table}

\vspace{-9pt}

\section{Discussion and Conclusions \label{sec:conclusion}}
Our results demonstrate that deconvolution with STARRED can significantly enhance the accuracy and precision of strong lens mass modelling. Specifically, the measurements of the Einstein radius, \thetae, and the axis ratio \qm reach sub-3\% precision, while the accuracy is better than 5\% for \thetae (except for SIE \#1) and 20\% for \qm. The configuration of SIE \#1 makes one of the images remain blended with the lens light; hence, it exhibits the smallest improvement in accuracy for \thetae and \qm. The improvements in accuracy and precision are driven by two key advantages of deconvolution: the enhanced resolution of source light and the ability to effectively mask lens light, which is critical for systems with small separations or blended components (e.g., SIE \#1 and \#2). These gains are especially valuable for ground-based surveys such as HSC, where atmospheric blurring and lower spatial resolution traditionally limit the fidelity of lens models.

The deconvolution process introduces non-physical artefacts, which cannot be perfectly modelled by standard Sersic profiles. As more complex profiles would not improve the fitting and would increase sources of bias, we mitigate this by focusing on the general shape of the arc. 
For this reason, we do not report the fitted source and
lens light parameters. The fitted light parameters are not expected to
be physically meaningful, and only the mass-model parameters, which we
show to be robustly recovered, are discussed. We evaluated this method using simulations with a known SIE mass parametrisation. While most real lenses conform to this model (e.g., Ref.~\cite{shajib19}), deviations are possible (e.g., Ref.~\cite{paic26}), potentially compromising the accuracy of the \thetae and \qm measurements. While flexible profiles such as the Elliptical Power Law could theoretically account for radial mass profile variations, the constraints on the radial slope often rely on magnification gradients. As the fitted source shape is altered, the measurement of a radial mass-profile slope, which relies on magnification gradients traced by the detailed shape of the arc, is compromised.  
Despite this limitation, when we apply our SIE+shear modelling approach to the two EPL mocks (with input slopes $\gamma = 1.87$ for EPL \#1 and $\gamma = 2.20$ for EPL \#2), we find that the precision and accuracy of the recovered \thetae and \qm are not significantly impacted relative to the SIE mocks (Figure~\ref{fig:thetae}), which is expected because \thetae is fixed by the tangential critical curve and is only weakly tied to the radial slope. This results supports the applicability of this approach beyond the SIE case tested here.
For the purpose of large-scale ground-based surveys such as SuGOHI, we advocate the SIE + shear model because it already enables robust scientific analyses across thousands of systems. We stress, however, that our seven-mock sample constitutes a proof of validity of the method rather than a population study.
We additionally tested the robustness of our approach against two methodological choices specific to the deconvolution step. First, varying the high-frequency regularisation strength $\lhf$ over two orders of magnitude left the recovered \thetae and \qm unchanged, confirming that these measurements were driven by the data rather than by the choice of regularisation (Appendix~\ref{app:lhf}). The external shear \gext, by contrast, was not recovered by our method: its magnitude shifted by 0.053 when the full noise covariance was used (Appendix~\ref{app:noisemap}), a change comparable to or larger than the true \gext values of our mocks (Table~\ref{tab:priors}), and its direction $\Psi_{\rm ext}$ was correspondingly unconstrained.

Second, since deconvolution correlates the pixel-level noise, we tested whether the diagonal noise covariance used in our fiducial likelihood optimistically improves the reported precision by repeating the fit with a full, propagated noise covariance (Appendix~\ref{app:noisemap}). We found that the sub-3\% precision was \thetae is preserved, once pixel-noise correlations were accounted for, at a computational cost roughly 10 times higher than the diagonal likelihood; given this cost and the modest impact on our results, we adopted the diagonal covariance as sufficient for the survey-scale goals of this work.

Looking ahead, we plan to extend this work by applying deconvolution on real HSC data to measure precise \thetae for all lenses in SuGOHI to improve, for instance, studies of the IMF \cite{sonnenfeld19} or galaxy evolution \cite{matsumoto10}. As the era of large-scale surveys such as Rubin--LSST dawns, methods such as ours will be critical for ensuring that ground-based observations remain a powerful tool for modelling. With open-source tools such as STARRED and Herculens, we invite the community to build upon these advances and maximise the scientific output of future strong lens system samples.

%%%%%%%%%%%%%%%%%%%%%%%%%%%%%%%%%%%%%%%%%%
\vspace{6pt}
\authorcontributions{Conceptualisation, E.P. and K.C.W.; methodology, E.P.; software, E.P.;  resources, A.M., A.T.J.; writing---original draft preparation, E.P.; writing---review and editing, K.C.W.; visualisation, E.P.; supervision, K.C.W.; funding acquisition, K.C.W. All authors have read and agreed to the published version of the manuscript.}

\funding{E.P. is supported by Japan Society for Promotion of Science (JSPS) KAKENHI Grant Number JP24H00221. K.C.W. is supported by JSPS KAKENHI Grant Numbers JP24K07089 and JP24H00221.}

\dataavailability{All simulated data come from \cite{jaelani24}. %MDPI: We encourage all authors of articles published in MDPI journals to share their research data. In this section, please provide details regarding where data supporting reported results can be found, including links to publicly archived datasets analyzed or generated during the study. Where no new data were created, or where data is unavailable due to privacy or ethical restrictions, a statement is still required. Suggested Data Availability Statements are available in section “MDPI Research Data Policies” at https://www.mdpi.com/ethics.
}

%\durcstatement{}

% Only for journal Nursing Reports
%\publicinvolvement{}
%
%% Only for journal Nursing Reports
%\guidelinesstandards{Please add a statement indicating which reporting guideline was used when drafting the report. For example, ``This manuscript was drafted against the XXX (the full name of reporting guidelines and citation) for XXX (type of research) research''. A complete list of reporting guidelines can be accessed via the equator network: \url{https://www.equator-network.org/}.}
%
%% Only for journal Nursing Reports
\useofartificialintelligence{Mistral AI tools were used in language editing, grammar and some parts of the code.}

\acknowledgments{The authors warmly thank Aymeric Galan for helpful discussions.}

\conflictsofinterest{The authors declare no conflict of interest. %MDPI: Declare conflicts of interest or state “The authors declare no conflict of interest.”.
} 

%%%%%%%%%%%%%%%%%%%%%%%%%%%%%%%%%%%%%%%%%%
%% Optional

%% Only for journal Encyclopedia
%\entrylink{The Link to this entry published on the encyclopedia platform.}

%\abbreviations{Abbreviations}{}

%%%%%%%%%%%%%%%%%%%%%%%%%%%%%%%%%%%%%%%%%%
%% Optional
\appendixtitles{yes} % Leave argument "no" if all appendix headings stay EMPTY (then no dot is printed after "Appendix A"). If the appendix sections contain a heading then change the argument to "yes".
\appendixstart
\appendix

\section[]{Tests Concerning the Deconvolved Cut-Out Noise Map \label{app:noisemap}}

We estimated the full pixel--pixel noise covariance matrix of the deconvolved cut-out via a Fisher-information analysis of the deconvolution objective $S$ (Equation~(\ref{eq:dec})), evaluated at the best-fit pixel grid $h(x)$, with no dependence on any lens-model parameter. We verified that the resulting correlation matrix is symmetric, positive definite, and bounded in $[-1,1]$ before propagating it into the lens-modelling likelihood, which we replaced with a multivariate Gaussian likelihood using this covariance.

\begin{figure}[H] 
    \includegraphics[width=0.98\textwidth]{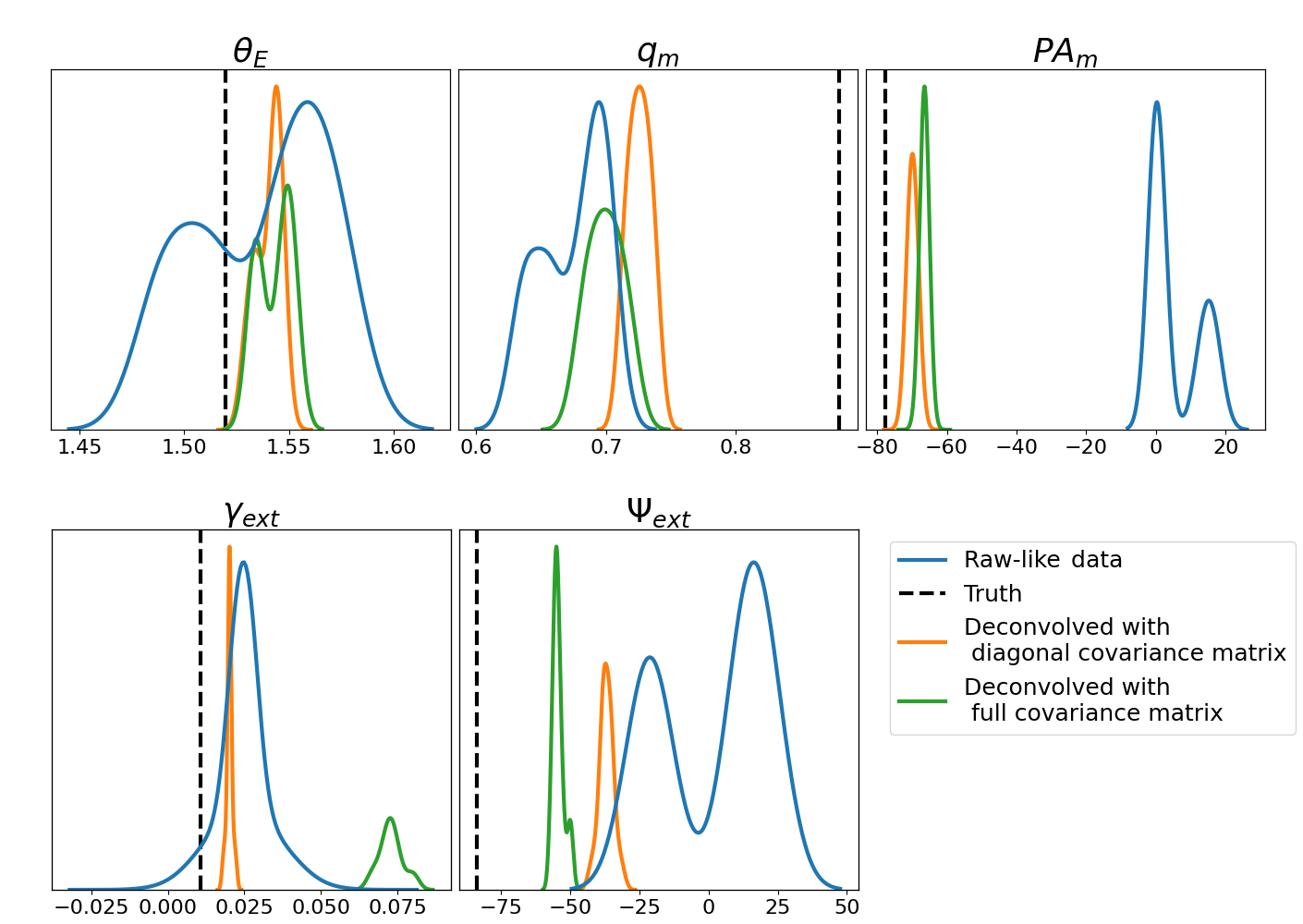}
    \caption{Measurement of mass parameters with raw-like data of the system SIE \#3 and deconvolved data using a diagonal or full covariance matrix.}
    \label{fig:correlated_noise_results}
\end{figure}

We repeated the lens-modelling fit for system SIE~\#3 using this correlated-noise likelihood; the results are shown in Figure~\ref{fig:correlated_noise_results}. The resulting precision on \thetae is 3\% with the full covariance compared to 2\% with the diagonal covariance for the same system. The corresponding accuracy is 1\% with full covariance versus 1\% with diagonal covariance. For \qm, precision is 12\% with full covariance versus 9\% with diagonal, and accuracy is 22\% with full covariance versus 18\% with diagonal. For $\gamma_{\rm ext}$, precision is 13\% with full covariance versus 5\% with diagonal, while the measurement is shifted by 0.053 away from the true value. As acknowledged in the body of the paper and in the previous paragraph, our method is well suited for measuring \thetae and \qm but not $\gamma_{\rm ext}$. We see that the precision and accuracy of the measurements of \thetae and \qm are slightly overestimated when a diagonal covariance matrix is used, but measurements with a full covariance matrix are still more precise and accurate than with the HSC-like data.

We also note that using the full covariance increases the computational cost of the lens-modelling fit from $\sim$1~h to $\sim$10~h per system, owing to the $\mathcal{O}(n^2)$ triangular solving required at each likelihood evaluation in place of the $\mathcal{O}(n)$ diagonal case. Given this cost and the results above, we advocate the diagonal covariance as sufficient to efficiently measure \thetae and \qm on large samples of lenses. 
Overall, this test confirms that the sub-3\% precision on \thetae
reported in this work is preserved, albeit only marginally, once
pixel-noise correlations introduced by deconvolution are propagated
into the lens-modelling likelihood.

\vspace{-4pt}

\section[]{Stability Across \boldmath{\lhf} \label{app:lhf}}
\begin{figure}[H]

    \includegraphics[width=0.98\textwidth]{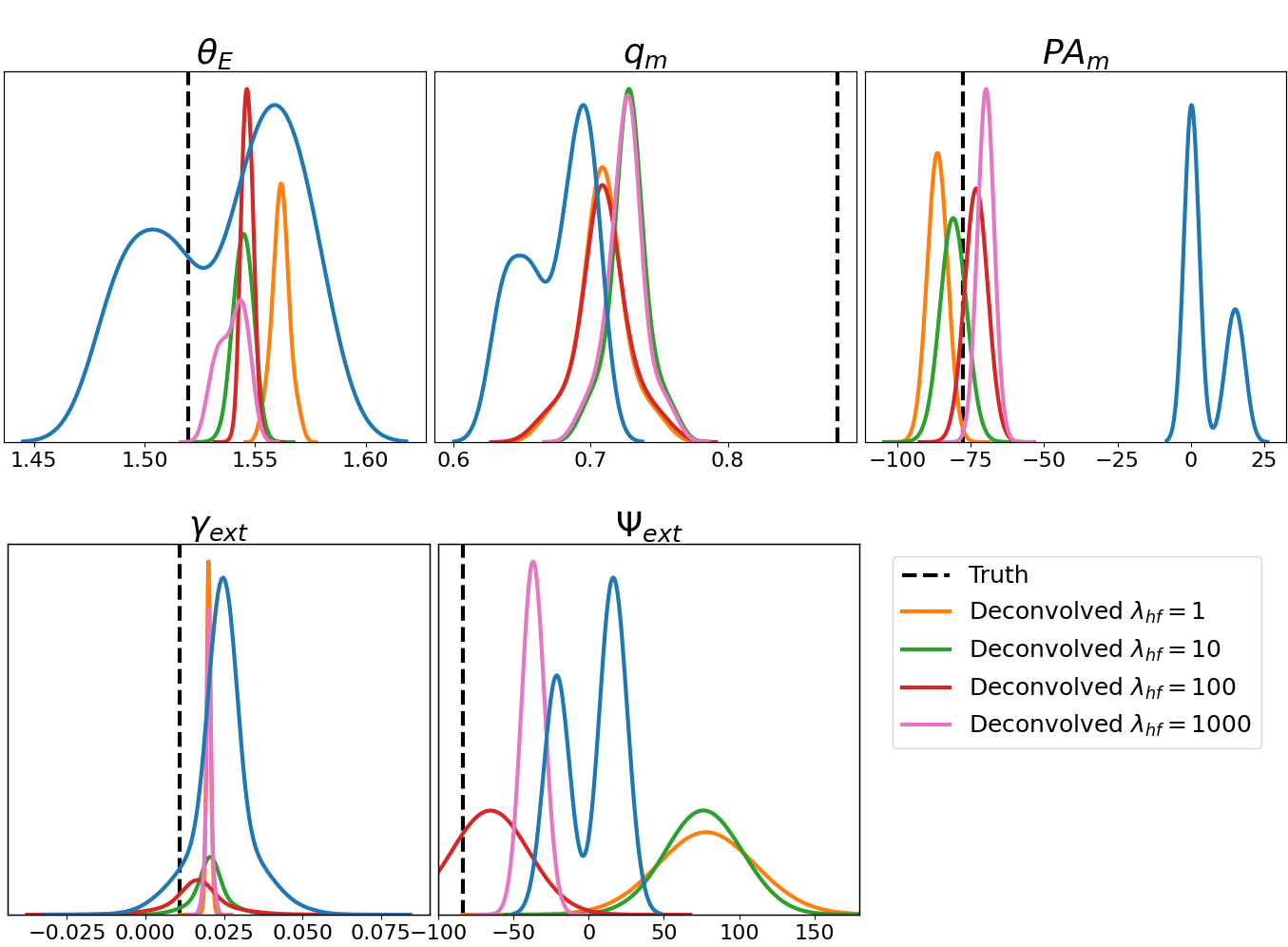}
    \caption{Measurement of mass parameters with HSC-like data of the system SIE \#3 and deconvolved data using \lhf $\in [1,10,100,1000]$. }
    \label{fig:lhfs}
\end{figure}

%\subsection[\appendixname~\thesubsection]{}
We performed an additional test on system SIE~\#3, varying $\lhf \in \{1, 10, 100, 1000\}$ while fixing $\lscale = 1$; the results are shown in Figure~\ref{fig:lhfs}. For \thetae and \qm, the posteriors remain tightly clustered and consistent across the entire range of \lhf values tested, and all are significantly closer to the truth than the HSC-like measurements. This demonstrates that the choice of \lhf does not significantly influence the recovered values for these parameters.

For the direction of the external shear, $\psi_{\rm ext}$, the posteriors' peaks shift within $200^\circ$ across the tested \lhf range. We attribute this to the low magnitude of $\gamma_{\rm ext}$, which results in its direction having a marginal impact on the modelling and thus being poorly constrained. This further supports our conclusion that our method is robust for measuring \thetae and \qm but not $\gamma_{\rm ext}$.

\vspace{-12pt}

%\section[\appendixname~\thesection]{}
%All appendix sections must be cited in the main text. In the appendices, Figures, Tables, etc. should be labeled, starting with ``A''---e.g., Figure A1, Figure A2, etc.

%%%%%%%%%%%%%%%%%%%%%%%%%%%%%%%%%%%%%%%%%%
%\isPreprints{}{% This command is only used for ``preprints''.
\begin{adjustwidth}{-\extralength}{0cm}

\reftitle{References}

\PublishersNote{}
%\isPreprints{}{% This command is only used for ``preprints''.
\end{adjustwidth}
\end{document}